\documentstyle[12pt]{article}
\newcommand{\erf}{\mbox{$\rm erf$}}
\newcommand{\bsigma}{\mbox{\boldmath $\sigma $}}
\newcommand{\bepsilon}{\mbox{\boldmath $\epsilon $}}

\begin{document}
\title{On the Origin of the Violation of Hara's Theorem for Conserved Current}
\author{
{P. \.{Z}enczykowski}$^*$\\
\\
{\em Dept. of Theor. Physics} \\
{\em Institute of Nuclear Physics}\\
{\em Radzikowskiego 152,
31-342 Krak\'ow, Poland}\\
}
\maketitle
\begin{abstract}
I elaborate on the argument that the
violation of Hara's theorem for conserved current requires that the current
is not sufficiently well localized. It is also stressed that whatever sign of
asymmetry is measured
in the $\Xi ^0 \to \Lambda \gamma $ decay,  
one of the following three
statements must be incorrect:
1) Hara's theorem is satisfied, 2) vector meson dominance is applicable to
weak radiative hyperon decays, and 3) basic structure of our 
quark-model description of
nuclear parity violation is correct.

\end{abstract}
$^*$ E-mail:zenczyko@solaris.ifj.edu.pl\\
\vskip 0.8in
\begin{center}REPORT \# 1827/PH, INP-Krak\'ow \end{center}
\newpage

\section{Introduction}

In 1964 Hara proved a theorem \cite{Hara}, according to which the 
parity-violating amplitude of the $\Sigma ^+ \rightarrow p \gamma $
decay should vanish in the limit of exact SU(3) symmetry.
The assumptions used in the proof were fundamental.
Over the years, however, there appeared several
theoretical, phenomenological and experimental indications that, 
despite the proof, Hara's theorem may be violated.
Quark model calculations of Kamal and Riazuddin \cite{KR},
VDM-prescription \cite{Zen89} and experiment \cite{Foucher,LZ}
provide such hints.
In particular only these
models that violate Hara's theorem provide a reasonably good
description of the overall body of experimental data on weak
radiative hyperon decays \cite{LZ}, as it stands now.

Obviously, if Hara's theorem is violated in Nature it follows that
at least one of its fundamental assumptions is not true.
This in turn means that some unorthodox and totally new physics 
must manifest itself in weak radiative hyperon decays (WRHD). Although
in general any non-orthodox physics should  
be avoided as long as possible,
the problem with WRHD is that we are on the verge of being
forced to accept it.
Namely, there exists a clean experimental way of distinguishing between the
orthodox and nonorthodox physics. The decisive measurable parameter is the
asymmetry of the $\Xi ^0 \to \Lambda \gamma $ decay.  Its absolute value
is expected to be large (of order  $0.7$) independently of 
the type of physics involved.  One may show that the sign of the asymmetry 
should be negative (positive) if physics  is orthodox (unorthodox).
Present experimental number is $+0.43\pm 0.44$, almost $3 \sigma$ away from the 
orthodox prediction. 
Of course the relevant experiment may have been performed or
analysed incorrectly. However, this is just one (though the most crucial) of
the hints against Hara's theorem.
Other hints, more theoretical in nature, 
are provided by the calculations in the naive quark model
\cite{KR} and by the VDM approach \cite{Zen89,LZ} in which VDM was
combined with our present knowledge
on parity violating weak coupling of vector mesons to nucleons \cite{DDH}.

There is a growing agreement that the calculation originally presented by
Kamal and Riazuddin (KR) is technically completely correct \cite{Azim,Holst}.
(However, there is no consensus as to the meaning of the KR result
\cite{Azim,Holst}.)
The VDM approach is based on two pillars: VDM itself and the DDH paper
on nuclear parity violation \cite{DDH}, in which
parity violating weak couplings of mesons to nucleons are discussed.
The DDH paper forms the
foundation of our present understanding of the whole subject of nuclear
parity violation, 
with the basis of the paper hardly to be questioned \cite{Holst}. 
Similarly, VDM has an extraordinary success record in low energy physics.
If Hara theorem is correct at least 
one of the above two pillars of the VDM approach to WRHD
must be incorrect.
This would be an important discovery in itself.

Given this situation, I think it is a timely problem
to pinpoint precisely what it is that might
lead to the violation of Hara's theorem.
Some conjectures in this connection were presented in \cite{LZ}
(and even earlier, see references cited therein).  These conjectures
pointed at the assumption of locality. In fact,
in a recent Comment \cite{Zen} it was shown that one can obtain violation of
Hara's theorem for conserved current provided the current is not sufficiently
well localized.  As proved in \cite{Zen}, the Hara's-theorem-violating 
contribution comes from $r=\infty $. 
However, as the example of the Reply \cite{Dmitra} to my Comment shows, 
the content and implications of the Comment are not always understood.
Therefore, in this paper I will try to shed some 
additional light on the problem.

Before I discuss the question of the implication of current (non)locality
on Hara's theorem I will show that the argument raised in \cite{Dmitra}
against the technical correctness of the KR calculation is logically incorrect.

After disposing of the argument against the technical correctness of the KR
calculation
I will present a simple example 
in which current conservation {\em alone}
does not ensure that Hara's theorem holds,
unless an {\em additional physical} assumption is made.

Then, I will proceed to discuss the main relevant point 
made in ref.\cite{Dmitra}.
In fact, ref.\cite{Dmitra} agrees with my standpoint that any violation
of Hara's theorem must result from a new phenomenon.  
However, identification of the origin of this phenomenon therein proposed 
 is mathematically incorrect.
This shall be proved below in several ways.

In the final remarks I will stress once again 
that the resolution of the whole issue (in favour of Hara's theorem or 
against it) can be settled once and forever by experiment,
that is by a mesurement of the asymmetry of the
$\Xi ^0 \rightarrow \Lambda \gamma$ decay.

\section{Conservation of the nonrelativistic current }

In ref.\cite{KR} Kamal and Riazuddin obtain gauge-invariant current-conserving
covariant amplitude. Ref.\cite{Dmitra} accepts correctness of their 
calculation up to this point. The claim of ref.\cite{Dmitra}
is that the authors of \cite{KR} incorrectly perform nonrelativistic
reduction thereby violating current conservation.
According to ref.\cite{Dmitra} this may be seen from Eq.(13) of ref.\cite{KR}
which is of the form $H_{PV}\propto \bepsilon \cdot 
({\bf \bsigma _1 \times \bsigma _2})$.
In this equation the current {\em seems} to be of the form
\begin{equation}
\label{KRcurrent}
{\bf J}=\bsigma _1 \times \bsigma _2
\end{equation}
and is not transverse as it should have been for a conserved current.

This claim is logically incorrect.
Eq.(13) of ref.\cite{KR} is obtained after {\em both}
performing the nonrelativistic reduction {\em and} choosing
the Coulomb gauge $\bepsilon \cdot {\bf \hat{q}} =0$
(${\bf \hat{q}}={\bf q}/|{\bf q}|$).
The origin of the lack of transversity of the "current" ${\bf J}$ in 
Eq.(\ref{KRcurrent}) is {\em not} 
the nonrelativistic reduction but the choice of 
Coulomb gauge $\bepsilon \cdot {\bf \hat{q}}=0$, i.e. the {\em restriction}
to transverse degrees of freedom only.
By choosing the Coulomb gauge we restrict the allowed $\bepsilon\ $ to
be transverse only.
It is then incorrect to replace $\bepsilon\ $ by (longitudinal) ${\bf \hat{q}}$.
In other words the correct form of the current-photon interaction
insisted upon in ref.\cite{Dmitra}, i.e.
\begin{equation}
\bepsilon \cdot (\bsigma _1 \times \bsigma _2 - {\bf \hat{q}}
[(\bsigma _1 \times \bsigma _2 )\cdot {\bf \hat{q}}])
\end{equation}
after choosing the Coulomb gauge 
$\bepsilon \cdot {\bf \hat{q}} = 0$ reduces to Eq.(13) of ref.\cite{KR}.  
Hence, from the form
$\bepsilon \cdot (\bsigma _1 \times \bsigma _2)$ obtained in ref.\cite{KR}
{\em after} choosing the Coulomb gauge one cannot conclude that 
the current is ${\bf J}=\bsigma _1 \times \bsigma _2$ and therefore that the 
nonrelativistic reduction was performed incorrectly.

Having proved that the argument against the KR calculation presented in
ref.\cite{Dmitra} 
is logically incorrect, we proceed to the issue of current (non)locality.

\section{A simple example}

Let us consider the well-known concept of partially conserved axial current (PCAC).
According to this idea the axial current is approximately conserved,
with its divergence proportional to the pion mass squared.
The weak axial current becomes divergenceless when the pion mass goes to zero,
a situation obtained in the quark model with massless quarks.
Thus, one may have a nonvanishing coupling of a vector boson to an axial
conserved current and a nonvanishing transverse electric dipole moment,
ie. violation of Hara's theorem. 

The price one has to pay to achieve this in the above 
example is the introduction of {\em massless} pions.
A massless pion corresponds to an interaction of
an infinite range - the pion may propagate to spatial infinity.  
Thus, vice versa, if one obtains
a nonvanishing transverse electric dipole moment 
in a gauge-invariant calculation (the KR case) this
suggests that the relevant current contains a piece that does not vanish at 
infinity sufficiently fast but {\em resembles} the pion contribution in the
example above. In other words one expects that something happens at
spatial infinity.

Of course, for Hara's theorem to be violated,
the mechanism of providing the necessary nonlocality must
be different from the particular one discussed above. 
After all, no massless hadrons exist. 
Consequently, current nonlocality would have to constitute an intrinsic feature
of baryons. It might result from baryon compositeness: it is known that
composite quantum states may exhibit nonlocal features. 
In this paper we will not pursue this line of thought any further since here
we are primarily interested in proving beyond any doubt
that nonlocality is crucial, but not
in discussing its deeper justification and implications.
Such a discussion will appear timely and desirable 
if new experiments confirm the positive sign
of the $\Xi ^0 \to \Lambda \gamma $ asymmetry.

Ref.\cite{Dmitra} accepts that
the current specified in ref.\cite{Zen} is conserved and that nonetheless 
it yields a nonzero value of the electric dipole moment in question.
However, it is alleged that this nonzero result originates
from $r=0$ (and not from spatial infinity).
In view of the example given above this claim should be 
suspected as incorrect. 
In fact its mathematical incorrectness can be proved.
Let us therefore see where the arguments of ref.\cite{Dmitra} break down.

\section{The origin of the nonzero contribution to the transverse 
electric dipole moment}
In ref.\cite{Zen} it is shown that for the current of the form
\begin{eqnarray}
{\bf J}_5^{\varepsilon }({\bf r})& =&
[\mbox{\boldmath{$\sigma $}}-
(\mbox{\boldmath{$\sigma $}}\cdot \hat{\bf r})\,\hat{\bf r}]
\, \delta _{\varepsilon }^3({\bf r})
+\frac{1}{2\pi r^2}\,
[\mbox{\boldmath{$\sigma $}}-
3(\mbox{\boldmath{$\sigma $}}\cdot \hat{\bf r})\, \hat{\bf r}]
\,\delta _{\varepsilon}(r) \nonumber \\
&& -
\frac{1}{4\pi r^3}\,
[\mbox{\boldmath{$\sigma $}}-
3(\mbox{\boldmath{$\sigma $}}\cdot \hat{\bf r})\, \hat{\bf r}]
\: {\rm erf} \left(\frac{r}{2\sqrt{\varepsilon }}\right)
\end{eqnarray}
where ${\rm erf} (x) = \frac{2}{\sqrt{\pi }} \int_{0}^{x}e^{-t^2}\,dt$
is the error function, 
$\hat{\bf r} = {\bf r}/r$, $r=|{\bf r}|$ and $\varepsilon \rightarrow 0$,
the transverse electric dipole moment is given by
\begin{equation}
\label{eq01}
T^{el}_{1M}=\lim_{\varepsilon \rightarrow 0}
\frac{iq}{2\pi \sqrt{2}} \int_0^{\infty} dr  
 \;\,
{\rm erf} \left( \frac{r}{2 \sqrt{\varepsilon }}\right)
\, j_1(qr)\; 
\int d\Omega_{\bf \hat{r}} \;
\mbox{\boldmath {$\sigma $}} \cdot {\bf \hat{r}} \,
\,Y_{1M}({\bf \hat{r}})
\end{equation}
and is nonzero.
The question is where does this nonzero result comes from.
Ref.\cite{Zen} (ref.\cite{Dmitra}) 
claim that the whole contribution is from
$r=\infty$ (respectively $r=0$).
We shall show that the claim of ref.\cite{Dmitra} is mathematically incorrect.

The Reply \cite{Dmitra} is based on the (true) equality (Eqs.(3,4) 
therein)
\begin{equation}
\label{eq1}
\alpha = \lim_{\epsilon \rightarrow 0} q \int_{0}^{\infty}
dr~j_1(qr)~\erf \left(\frac{r}{2\sqrt{\epsilon}} \right)=
\left( \frac{2}{q}\right) \int_{0}^{\infty}dz~j_0(z)~\delta 
\left(\frac{z}{q}\right) 
\end{equation}
in which the left-hand side (l.h.s.)
 is the original integral appearing in the expression for the electric dipole
 moment, from which it
was concluded in ref.\cite{Zen} that violation of Hara's theorem originates
from $r=\infty$. 

The Reply \cite{Dmitra} further claims that 
as one has to perform the integral
first, and only then take the limit $\epsilon \rightarrow 0$, it can be seen
from the right-hand side of Eq.(\ref{eq1}) that in the limit
$\epsilon \rightarrow 0$ the integral on the left-hand
side receives all its contribution from the point $r=0$.

That this claim is mathematically incorrect can be seen in many ways.
We shall deal with the integral on the left-hand side {\em directly} since
equality of definite integrals does not mean that the integrands are
identical.
In particular integration by parts used to arrive at the r.h.s. of 
Eq.(\ref{eq1}) may change the region from which the value of the integral
comes as it should be obvious from the following example:
\begin{eqnarray}
\int _0^{\infty}dx~\exp (-x)~\theta (x-\epsilon)&=&\nonumber \\
=-\exp (-x)~\theta (x-\epsilon)|_0^{\infty}&+&
\int_0^{\infty}dx~\exp (-x)~\delta (x-\epsilon)\nonumber \\
&=&\int_0^{\infty}dx~\exp (-x)~ \delta (x-\epsilon)
\end{eqnarray}
Clearly, the integral on the l.h.s. of Eq.(6)
does not receive all its contribution
from the point $x=\epsilon$ while the r.h.s. does.
Let us therefore concentrate on the l.h.s of Eq.(\ref{eq1})
since it is the integrand on the l.h.s. which has a physical meaning.

a) {\em Mathematical proof}

For {\em any finite} $\epsilon $ the integrand on the l.h.s. of Eq.(\ref{eq1})
vanishes for $r=0$ since $j_1(0)=\erf (0/(2\sqrt{\epsilon}))=0$.  Consequently,
already the most naive argument seems to show that
the point $r=0$ does not contribute in the limit $\epsilon \rightarrow 0$ at
all.
Should one be concerned with the neighbourhood of the point $r=0$, we notice
that both functions $j_1(qr)$ and $\erf (r/(2\sqrt{\epsilon}))$ are
{\em bounded} for any $q$, $r$, $\epsilon$ of interest.  
Consequently, the integrand 
on the left-hand side of Eq.(\ref{eq1}) is bounded by 
$\max _{0 \le z \le \infty}~j_1(z) \equiv M < \infty $.  
Hence, the contribution
from any interval $[0,\Delta]$, ($0\leq \Delta \ll 1 $) is bounded by
$q\int_0^{\Delta }dr~M \approx q \Delta M$ and vanishes when 
$q\Delta \rightarrow 0$. 
From the mathematical point of view the incorrectness 
of ref.\cite{Dmitra} is thus proved.

For further clarification, however, the following two points may be consulted.
Point b) below provides simple and intuitive visual demonstration 
of what happens on the l.h.s.
of Eq.(\ref{eq1}) in the limit $\epsilon \rightarrow 0$ .
In point c) the integral is actually performed {\em before} taking the limit
$\epsilon \rightarrow 0$, the procedure considered 
in ref.\cite{Dmitra} to be correct.

b) {\em Intuitive "proof"}

The integral on the left of Eq.(\ref{eq1}) can be evaluated for any 
$\epsilon $ (formula 2.12.49.6 in ref.\cite{Rus}) and one obtains
\begin{equation}
\label{eq2}
q \int_{0}^{\infty}
dr~j_1(qr)~\erf \left(\frac{r}{2\sqrt{\epsilon}} \right)=
\frac{\sqrt{\pi}}{2q}~\frac{1}{\sqrt{\epsilon}}~\erf (q\sqrt{\epsilon})
\end{equation}
which for small $q\sqrt{\epsilon}$ is equal to
\begin{equation}
\label{eq3}
1-\frac{q^2\epsilon}{3}+O((q^2\epsilon)^2)
\end{equation}
This approach to $1$ from below (when $q^2 \epsilon \rightarrow 0$) 
can be seen from a series of plots
shown in Fig.1.

In Fig.1 one can see that for small $q\sqrt{\epsilon}$ the integrand in
Eq.(\ref{eq1}) differs significantly from $j_1(qr)$ only for very small
$qr < q\Delta$, where the integrand is {\em smaller} than $j_1(qr)$.
It is also seen that in the limit $q\sqrt{\epsilon} \rightarrow 0$
the contribution from the region of small $qr$ {\em grows}
(thus the whole integral grows in agreement with Eq.(\ref{eq3}))
but {\em never} exceeds the integral $\int_0^{\Delta}q~dr~j_1(qr)$.
It is intuitively
obvious that the latter integral is smaller than $j_1(q\Delta)\cdot q\Delta$
and cannot yield the value $1$ in Eq.(\ref{eq3}) for $\Delta \rightarrow 0$!
For more details consult point (c2) below.


c) {\em Doing integrals first}

Should one be not satisfied for any reasons 
with the above two arguments, and 
insist
that one has to perform the integral first, an appropriate rigorous proof of 
mathematical incorrectness of ref.\cite{Dmitra} follows.  
In this proof the integral is performed {\em before}
taking the limit $\epsilon \rightarrow 0$, as argued in ref.\cite{Dmitra} 
to be the only correct procedure.

Let us divide the integral on the left-hand side of Eq.(\ref{eq1}) into
two contributions:
\begin{equation}
\label{eq4}
\lim_{\epsilon \rightarrow 0} \left[
\int_{0}^{\Delta}
dr~q~j_1(qr)~\erf \left(\frac{r}{2\sqrt{\epsilon}} \right)+
\int_{\Delta}^{\infty}
dr~q~j_1(qr)~\erf \left(\frac{r}{2\sqrt{\epsilon}} \right)
\right]
\end{equation}
where $\Delta $ is finite, but otherwise arbitrary: $0<\Delta<\infty$.

According to ref.\cite{Dmitra}, the whole
contribution to the integral on the left-hand side of Eq.(\ref{eq1}) comes
from the point $r=0$ when the limit $\epsilon \rightarrow 0$ is taken
{\em after} evaluating the integral.
Hence, the whole contribution to the left-hand side of Eq.(\ref{eq1})
should come from the first term in Eq.(\ref{eq4}), i.e. from
\begin{equation}
\label{eq5}
f_{[0,\Delta]}(q,\epsilon) \equiv 
\int_{0}^{\Delta}
dr~q~j_1(qr)~\erf \left(\frac{r}{2\sqrt{\epsilon}} \right)
\end{equation}
when the limit $\epsilon \rightarrow 0$ is taken {\em after} evaluating
the integral.

c1) Let us therefore estimate the integral $f_{[0,\Delta]}(q,\epsilon)$.
Integrating by parts we obtain
\begin{eqnarray}
f_{[0,\Delta]}(q,\epsilon)&=&-\frac{1}{q}j_0(q\Delta)\frac{2}{\sqrt{\pi}}
\int_0^{\Delta/(2\sqrt{\epsilon})}\exp (-t^2)~dt \nonumber\\
&&+\frac{1}{q}j_0(q\cdot 0)\frac{2}{\sqrt{\pi}}
\int_0^{0/(2\sqrt{\epsilon})}\exp (-t^2)~dt      \nonumber\\
&&+\frac{2}{\sqrt{\pi}}\frac{1}{2\sqrt{\epsilon}}\frac{1}{q}
\int_0^{\Delta}dr~j_0(qr)~\exp (-r^2/(4 \epsilon))
\end{eqnarray}
Since we take the limit $\epsilon \rightarrow 0$ only {\em after} evaluating
the integral, the second term above vanishes.
Thus
\begin{equation}
\label{eq6}
f_{[0,\Delta]}(q,\epsilon)=\frac{1}{q}\frac{2}{\sqrt{\pi}}
\int_0^{\Delta/(2\sqrt{\epsilon})}dt~\exp (-t^2)~
(j_0(q\cdot 2\sqrt{\epsilon }t)-j_0(q\Delta))
\end{equation}
Consequently
\begin{equation}
\label{bounding}
|f_{[0,\Delta]}(q,\epsilon )|\le \frac{1}{q}\frac{2}{\sqrt{\pi}}
\int_0^{\Delta/(2\sqrt{\epsilon})}dt~\exp (-t^2)~
|j_0(q\cdot 2\sqrt{\epsilon }t)-j_0(q\Delta)|
\end{equation}
We are ultimately interested in the limit $q \rightarrow 0$.  
Hence, let us
take $q\Delta \ll 1$. 
This may be assumed for {\em any} finite $\Delta$.
Since $0\le 2\sqrt{\epsilon}t\le \Delta$, and the function
$j_0(z)$ is monotonically decreasing for $z \ll 1$ it follows that
\begin{equation}
|j_0(q2\sqrt{\epsilon}t)-j_0(q\Delta)|\le |j_0(0)-j_0(q\Delta)|
\end{equation}
Hence, for $q\ll 1/\Delta$ we have
\begin{eqnarray}
|f_{[0,\Delta]}(q,\epsilon)|&\le&
\frac{1}{q}\frac{2}{\sqrt{\pi}}\int_0^{\Delta/(2\sqrt{\epsilon})}dt~
\exp (-t^2) |j_0(0)-j_0(q\Delta)|\nonumber \\
&\le&
\frac{1}{q}\frac{2}{\sqrt{\pi}}\int_0^{\infty}dt~
\exp (-t^2) |j_0(0)-j_0(q\Delta)|\nonumber \\
&=&\Delta \left|\frac{j_0(q\Delta)-j_0(0)}{q\Delta}\right|
\end{eqnarray}
For finite $\Delta $, in the limit $q\rightarrow 0$, the factor under the sign
of modulus is the definition of the derivative of $j_0$ at $0$, i.e.
\begin{equation}
\lim_{q \rightarrow 0}|f_{[0,\Delta]}(q,\epsilon)|\le \Delta |j_1(0)|
\end{equation}
Since $j_1(0)=0$ we conclude that for {\em any finite}
$\Delta $ one has $\lim_{q\rightarrow 0}|f_{[0,\Delta]}(q,\epsilon)|=0$,
and that this occurs for any finite $\epsilon$.
We now take the limit $\epsilon \rightarrow 0$ and obviously obtain
$\lim_{\epsilon \rightarrow 0} 
(\lim_{q\rightarrow 0}|f_{[0,\Delta]}(q,\epsilon)|)=0$.
This directly contradicts the claim of ref.\cite{Dmitra}.
It is also seen that only for $\Delta = \infty $ the above proof does not
go through because then $q\Delta$ is $\infty$
for any finite $q$, and
$|j_0(0)-j_0(q\Delta)|=|j_0(0)-j_0(\infty)|=|j_0(0)|=1$.
Thus, since for any finite $\Delta$ the contribution to the first term 
in Eq.(\ref{eq4}) is $0$ in the limit of $q \rightarrow 0$, the whole
contribution must come from the second term in Eq.(\ref{eq4}).
Since $\Delta$ is arbitrary, the contribution comes from $r=\infty$.
This can be checked by a direct evaluation of the second term in Eq.(\ref{eq4})
for {\em any finite} $\Delta$.

c2) Should someone be not convinced 
by the procedure of bounding the integrand
in Eq.(\ref{bounding}), one can perform the integral in Eq.(\ref{eq5}) directly.
Denoting $\delta = q\Delta$, $\epsilon ' = q \sqrt{\epsilon} $ we have
\begin{eqnarray}
f_{[0,\Delta]}(q,\epsilon)&=&\int_0^{\delta}dz~j_1(z)~\erf 
\left(\frac{z}{2\epsilon '}\right)= \nonumber\\
=-j_0(\delta) \erf \left(\frac{\delta}{2\epsilon '} \right)&+
&\int_0^{\delta}dz~j_0(z)\cdot \frac{2}{\sqrt{\pi}} 
\exp \left(-\frac{z^2}{4 \epsilon '^2}\right)\frac{1}{2\epsilon '}
\end{eqnarray}
For small $\epsilon '$ the second term on the r.h.s. above receives
contributions from small $z$ only.  Therefore we may expand $j_0(z)$
around $z=0$:
\begin{equation}
j_0(z)\approx 1-\frac{1}{6}z^2+...
\end{equation}
and perform the integrations.
We obtain
\begin{eqnarray}
\label{integrate}
\frac{2}{\sqrt{\pi}}\cdot \frac{1}{2\epsilon '}
\int_0^{\delta}dz~(1-z^2/6)\exp (-z^2/(4\epsilon '^2))=&&\nonumber\\
=\frac{2}{\sqrt{\pi}}\int_0^{\delta/(2 \epsilon')}dt~\exp(-t^2)-
\frac{1}{6}\cdot \frac{2}{\sqrt{\pi}}(2\epsilon')^2
\int_0^{\delta/(2\epsilon ')}dt~t^2~\exp (-t^2)&&
\end{eqnarray}
The integral in the second term in Eq.(\ref{integrate}) may be
evaluated as
\begin{eqnarray}
\label{secondterm}
\frac{2}{\sqrt{\pi}}\left[ 
-\frac{d}{d\lambda}\int _0^{\delta/(2\epsilon ')}dt~\exp (-\lambda t^2)
\right]_{\lambda =1}=&&\nonumber\\
-\frac{d}{d\lambda}\left[
\lambda ^{-1/2} \erf \left( \frac{\delta \lambda ^{1/2}}{2\epsilon '} \right)
\right]_{\lambda =1}=&&\nonumber\\
=-\frac{1}{2} \erf \left( \frac{\delta }{2 \epsilon '}\right)+
\frac{2}{\sqrt{\pi}}\frac{\delta }{4 \epsilon '} 
\exp (-\delta ^2/(4 \epsilon'^2))
\end{eqnarray}
Putting together Eqs.(13-\ref{secondterm})
one obtains
\begin{equation}
\label{final}
f_{[0,\Delta]}(q,\epsilon )=
(1-j_0(\delta )-\frac{\epsilon '^2}{3}) \erf (\delta /(2\epsilon '))+
\frac{\epsilon '^2}{3} \frac{2}{\sqrt{\pi}}\frac{\delta }{2\epsilon '}
\exp (-\delta^2 /(4 \epsilon'^2))
\end{equation}
We now recall that $\delta/\epsilon ' = \Delta /\sqrt{\epsilon}$
and that we are interested in the limit $\epsilon \rightarrow 0$ for
any finite $\Delta $.
For very large (but finite) $\delta $ and small $\epsilon '$
we have $j_0(\delta) \approx 0$, $\erf (\delta/(2\epsilon ')) \approx 1$,
and
\begin{equation}
\frac{\delta}{2 \epsilon '} \exp (-\delta/(4 \epsilon '^2))\approx 0.
\end{equation}
Eq.(\ref{final}) reduces then to
\begin{equation}
f_{[0,\Delta]}(q,\epsilon) \approx 1-\epsilon'^2/3
\end{equation}
approaching 1 from below
in agreement with Eq.(\ref{eq3}) and Fig. 1.

For $\epsilon \rightarrow 0$ and fixed $\Delta$ one obtains from 
Eq.(\ref{final})
\begin{equation}
\label{limes}
\lim _{\epsilon \rightarrow 0}f_{[0,\Delta]}(q,\epsilon)=
1-j_0(q\Delta)
\end{equation}
Clearly, the contribution to the integral in Eq.(\ref{eq1}) coming from
the interval $[0,\Delta]$ is small and goes to zero when 
$q\Delta \rightarrow 0$.
Thus, for {\em any finite} $\Delta$, in the limit $q \rightarrow 0$
the contribution to the integral in Eq.(\ref{eq1}) comes entirely from
the second term in Eq.(\ref{eq1}).  Since $\Delta$ is arbitrary,
the contribution comes from $r=\infty $.

\section{Final remarks}

In summary, 
violation of Hara's theorem may occur for conserved current as shown in
ref.\cite{Zen}.
One has to pay a price, though: the price is the lack of sufficient 
localizability of the current. This connection to the {\em physical} issue
of locality has been already suggested in \cite{LZ}.
Thus, violation of Hara's theorem
would require a highly non-orthodox resolution. 
Whether this is a physically reasonable option constitutes a completely separate
question.
However, one should remember that what is "physically reasonable" 
is determined by {\em experiment} and not by our preconceived ideas
about what the world looks like. 
After all, {\em all} our fundamental ideas are abstracted from experiment.
They do not live their own independent life and must be modified if
experiment proves their deficiencies.

In general, we should try to avoid non-orthodox physics as long as we can.
The problem is, however, that there are various
theoretical, phenomenological, experimental and even philosophical
hints that, 
despite expectations based on standard views, Hara's theorem may be violated.
It is therefore important to ask and answer the question whether one can
provide a single and clearcut test, the results of which would unambiguously  
resolve the issue.

In fact, as already mentioned in the introduction, 
such a test has been pointed out in \cite{LZ}
(see also \cite{APPB,Zen2000}).  
It was shown there
that the issue can be experimentally settled by measuring the asymmetry
of the $\Xi ^0 \rightarrow \Lambda \gamma$ decay. 
The sign of this asymmetry is strongly correlated with the answer to the
question of the violation of Hara's theorem in $\Sigma ^+\rightarrow p \gamma$. 
In Hara's-theorem-satisfying
models this asymmetry is negative and around $-0.7$.
On the contrary, in Hara's-theorem-violating models this asymmetry is
positive and of the same absolute size, (ie. it is around $+0.7$).
Present data is $+0.43 \pm 0.44$. The KTeV experiment at Fermilab
has 1000 events
of $\Xi ^0 \rightarrow \Lambda \gamma$ \cite{Ramberg99}. 
These data are being analysed.
Thus, the question of the violation of Hara's theorem should be
experimentally settled soon.

If the results of the KTeV experiment (and those of an even higher statistics
experiment being performed by the NA48 collaboration at CERN \cite{Koch}) 
confirm large positive asymmetry for
the $\Xi ^0 \rightarrow \Lambda \gamma$ decay, one should start to discuss
the possible deeper  physical meaning of the violation of Hara's theorem.
I tried to refrain from such a discussion so far.

On the other hand, if the asymmetry in the $\Xi ^0 \rightarrow \Lambda
\gamma $ decay is negative, one must conclude that Hara's theorem holds
in Nature.  In this case, however, it follows that either vector meson
dominance is inapplicable to weak radiative hyperon decays or our present
understanding of nuclear parity violation (ref.\cite{DDH}) is incorrect. 

In conclusion, whatever sign of asymmetry is measured in
the  $\Xi ^0 \rightarrow \Lambda \gamma $ decay, something well accepted
will have to be discarded.

\section{Acknowledgements}

I would like to thank A. Horzela for providing reference \cite{Rus} and 
J. Lach and A. Horzela for discussions 
regarding the presentation of the argument.
Comments on the presentation of the material of this paper,
received from V. Dmitrasinovic prior to paper's dissemination,
are also gratefully acknowledged.


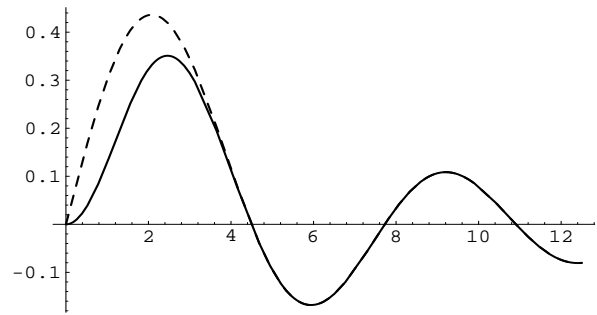
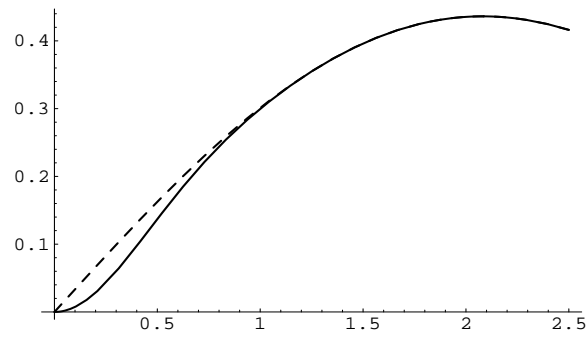
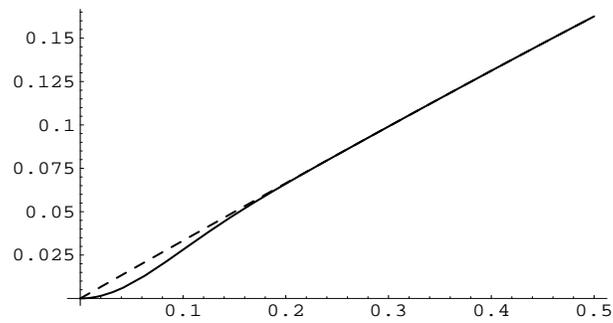
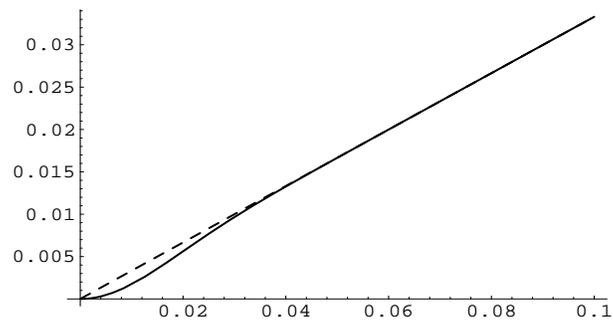
\end{document}